\documentstyle[12pt,A4wide]{article}
\begin{document}
\rightline{April the 8th, 1992}
{\begin{center} {\Large Semi-Classical Quantization of \\
\vspace{0.3cm}
                        the Many-Anyon System} \\

\vspace{2cm}
{\large Fabrizio Illuminati\footnote{Bitnet
address: illuminati@padova.infn.it}}  \\
\vspace{0.3cm}
{\it Dipartimento di Fisica ``Galileo Galilei'', Universit{\`a} \\
di Padova, Via F.Marzolo 8, Padova 35131, Italia.} \end{center}}

\vspace{2cm}
{\begin{center} \large \bf Abstract \end{center}}

We discuss the problem of $N$ anyons in harmonic well, and
derive the semi-classical energy spectrum as an exactly
solvable limit of the many-anyon Hamiltonian. The relevance
of our result to the solution of the anyon-gas model is discussed.

\vspace{0.4cm}

\vfill
DFPD 92/TH/26 \hfill May 1992
\newpage
{\large \bf 1. Introduction}

\vspace{0.4cm}

The possibility of arbitrary statistics continously
interpolating between the Bose and the Fermi case in
space dimensions less than three was suggested some
years ago in a beautiful and pioneering paper [1] by
Leinaas and Myrheim. Their analysis showed that
the topology of a non-simply connected configuration
space affects the behaviour of a quantum system of identical
particles exactly like a multiply connected physical space
determines effects of the Aharonov-Bohm type [2]. A physical
model of particles with arbitrary, or fractional, statistics
was put forward by Wilczek [3], who baptized such objects
as ``anyons". Anyons are two-dimensional particles endowed
with fictitious statistical charge and magnetic flux; their
wavefunctions will in general be multivalued, due to the
Aharonov-Bohm phases that the anyons acquire in winding
around each other on the plane of motion. The relevance of
the anyon model has been suggested for such important phenomena
as the fractional quantum Hall effect and high temperature
superconductivity [4], [5]. It is then crucial to acquire the
deepest possible knowledge of the properties of the general
$N$-body anyon system. Unfortunately, despite growing effort in
the last few years, only the two-anyon problem has been exactly
solved by Leinaas and Myrheim in their original paper [1]. For
the three-anyon problem Wu has obtained a special set of exact
solutions [6] which have been recently generalized to the
arbitrary $N$-anyon system in a magnetic field by Dunne {\em
et al.} [7]. Here we will not pursue this direction, but instead
concentrate on a unified approximate treatment of the many-anyon
problem in the framework of the semi-classical quantization.

\vspace{1cm}

{\large \bf 2. Semi-classical approximation}

\vspace{0.5cm}

In a previous paper [8], a semi-classical quantization method was
put forward to study the two- and three-anyon systems. It was
shown that the method yields the exact energy spectrum for two
anyons. In the three-anyon case it gives a spectrum linear in the
statistical parameter and in good agreement with the numerical one
obtained by Sporre {\em et al.} [9], except for a slight
nonlinearity in some of the numerical levels.
Here we briefly review the main points of the argument. Consider a
system of two non-interacting anyons in an external harmonic
oscillator potential of frequency $2\pi\omega$; the classical
Hamiltonian for the relative motion in polar coordinates $(r,\phi)$
reads

\begin{equation}
H = \frac{1}{2}p_{r}^{2} +\frac{1}{2}\omega^{2}r^{2} +
\frac{1}{2r^{2}}(p_{\phi} - \alpha\hbar)^{2},
\end{equation}

\noindent where $\alpha$ is the statistical parameter; Bose
statistics is recovered for $\alpha = 0$ and Fermi statistics
for $\alpha =1$. From the classical Hamiltonian
one can build the classical partition function

\begin{equation}
Z_{class} = \frac{1}{(
2\pi\hbar)^{2}}{\int}drdp_{r}d{\phi}dp_{\phi}
e^{{-\beta}H}.
\end{equation}

\noindent The semi-classical strategy amounts now to a
discretization of the relevant
degrees of freedom to obtain a partition function expressed
as a sum of
Boltzmann factors. By identifying the energy eigenvalues
in the exponents
one can obtain the energy spectrum of the
system. The essential quantum
behaviour of the anyon model is contained
in the angular degrees of freedom.
Then, letting $p_{\phi}\rightarrow{\hbar}m$ and
${\int}dp_{\phi}\rightarrow\hbar\sum_{m}$, $m$ being
the angular momentum quantum number, the semi-classical
partition function reads

\begin{equation}
Z_{s.-c.} = \frac{1}{2\beta\hbar\omega}
\sum_{m}\exp{(\beta\hbar\omega{|m-\alpha|})}.
\end{equation}

The semi-classical quantization is most significant in the
small quanta of action regime, i.e. when $\beta\hbar\omega\ll1$.
In this limit it is correct to approximate $\beta\hbar\omega$ by
$\sinh{\beta\hbar\omega}$
in the denominator of (3). By use of a power series representation
for
$(2\sinh{\beta\hbar\omega})^{-1}$ one can finally put $Z_{s.-c.}$
in the desired form

\begin{equation}
Z_{\it s.-c.} =
\sum_{m,n}\exp{(-\beta(2n+1+|m-\alpha|)\hbar\omega)},
\end{equation}

\noindent and read off the energy eigenvalues

\begin{equation}
E_{m,n} = (2n + 1 + |m - \alpha|)\hbar\omega,
\end{equation}

\noindent which turn out to be the exact ones [1].
Extension of the method to systems
of more than two anyons meets the obstacle of the
partition function being
not factorizable in a product of one-dimensonal integrals.
To by-pass the problem, in [8] some further approximations
were made. The result was a semi-classical partition function
for the relative motion of three anyons expressed as the
product of two partition functions for
the relative motion of two anyons, with statistical parameters
$\alpha$ and $2\alpha$ respectively.
Going again through the semi-classical procedure sketched above,
the energy
spectrum for the relative motion of three anyons was found to be

\begin{equation}
E = (2n_{1} + 1 + 2n_{2} + 1 + |m_{1} - \alpha| + |m_{2} -
2\alpha|)\hbar\omega).
\end{equation}

The spectrum (6) is drawn in Fig.1. It is
again linear in $\alpha$, and the energy levels fall into two
classes, one with slopes ${\pm}1$, and the other with
slopes ${\pm}3$.
Remarkably, this behaviour agrees with the numerical solution
of the
three-anyon problem [9], although the semi-classical levels
with slopes
${\pm}1$ appear to be slightly curved in the numerical
solution.
Furthermore, the semi-classical levels with slopes ${\pm}3$
are equal to
the exact eigenvalues obtained from the special solutions
of Wu [6].
Thus the energy relations (5) and (6) seem to
capture some of the essential features of the few-anyons
physics. In this
work we shall provide a deeper understanding of this fact,
and a precise
setting of the semi-classical quantization in terms of a
limit of
separability of the many-anyon Lagrangian. This connection
explains
the nature of the approximations made in deriving the
semi-classical energy
relation (6) and allows its immediate generalization to
systems of
arbitrarily many anyons. This semi-classical $N$-anyon energy
spectrum and
the numerical one are compared for the case $N=4$ and are
again shown to be in good agreement.
\vspace{0.8cm}

{\large \bf 3. Separation of the many-anyon problem}

\vspace{0.5cm}

We consider again eq. (6). After a moment's thought one
realizes that it
is the sum of the energy spectra of two internal
oscillators, one with
statistical parameter $\alpha$, the other with statistical
parameter $2\alpha$.
This suggests that the approximation made in [8] to
factorize the partition function might
simply amount to replace the exact relative hamiltonian with
a separable one.
This observation would also explain why the semi-classical
spectrum is
exact in the case of two anyons, since the exact two-anyon
lagrangian is
already separated. We first recall that a non-relativistic
system of $N$ anyons
in an external harmonic oscillator potential is described,
in cartesian coordinates, by the Lagrangian

\begin{equation}
L = \frac{1}{2}\sum_{i=1}^{N}({\dot{\bf r}}_{i}^{2}
- {\omega}^{2}{\bf r}_{i}^{2}) +
\alpha\hbar\sum_{i<j}{\dot{\phi}}_{ij},
\end{equation}

\noindent where ${\bf r}_{i}=(x_{i},y_{i})$, and the azimuthal
angle $\phi_{ij}$ is defined by

\begin{equation}
\phi_{ij} = \arctan{\frac{y_{j} - y_{i}}{x_{j} - x_{i}}}.
\end{equation}

The separation of the center of mass motion can be achieved
through the change of variables

\begin{equation}
{\bf r}_{1},\, {\bf r}_{2},\, \ldots,\,
{\bf r}_{\scriptscriptstyle{N}}\,
\,\longrightarrow \, \, {\bf R},\,
{\mbox{\boldmath $\rho$}}_{1},\,
{\mbox{\boldmath $\rho$}}_{2},\, \ldots,\,
{\mbox{\boldmath $\rho$}}_{\scriptscriptstyle{N-1}},
\end{equation}

\noindent with the center of mass coordinate $\bf R$ and the
$N-1$ Jacobi coordinates
$\{ {\mbox{\boldmath $\rho$}}_{k}\}$ defined as

\begin{eqnarray}
{\bf R} & = & \frac{1}{\sqrt{N}}({\bf r}_{1} +
{\bf r}_{2} + \cdots +
   {\bf r}_{\scriptscriptstyle{N}}), \nonumber \\
{\mbox{\boldmath $\rho$}}_{1} & = &
\frac{1}{\sqrt{2}}({\bf r}_{1} -
{\bf r}_{2}), \nonumber \\
{\mbox{\boldmath $\rho$}}_{2} & = &
\frac{1}{\sqrt{6}}({\bf r}_{1} + {\bf r}_{2} -
                     2{\bf r}_{3}), \nonumber \\
               & \vdots & \nonumber \\
{\mbox{\boldmath $\rho$}}_{\scriptscriptstyle{N} -
1} & = & \frac{1}{\sqrt{N
(N-1)}}({\bf r}_{1} + {\bf r}_{2} + \cdots +
{\bf r}_{\scriptscriptstyle{N} -
1} - (N-1){\bf r}_{\scriptscriptstyle{N}}).
\end{eqnarray}

\noindent For the Jacobi coordinates the associated
angles relative to the {\bf x}-axis are

\begin{eqnarray}
\phi_{k} & = & \arctan{\frac{y_{1} + y_{2} + \cdots
+ y_{k} - ky_{k+1}}{x_{1} + x_{2} + \cdots + x_{k} -
kx_{k+1}}}, \nonumber \\
            k & = & 1,2,\ldots ,N-1.
\end{eqnarray}

\noindent Then, neglecting the center of mass motion and
expressing the Jacobi
coordinates in the polar form $\{\rho_{k},\phi_{k}\}$ (with
$\rho_{k} =
\sqrt{{\mbox{\boldmath $\rho$}}_{k}\cdot{\mbox{\boldmath
$\rho$}}_{k}}$),
eq.(7) becomes for the relative motion

\begin{equation}
L_{rel} = \frac{1}{2}\sum_{k=1}^{N-1}({\dot{\rho}}_{k}^{2} +
{\rho}_{k}^{2}{\dot{\phi}}_{k}^{2}) -
\frac{1}{2}\sum_{k=1}^{N-1}{\omega}^{2}{\rho}_{k}^{2} +
\alpha\hbar\sum_{i<k}{\dot{\phi}}_{ik}.
\end{equation}

We now investigate the relation between the azimuthal
angles $\phi_{ik}$ and
the polar Jacobi angles $\phi_{k}$; consider first the
situation in which
the $N$ anyons are arranged in ``hierarchical clusters", i.e.

\begin{equation}
\rho_{1} \ll \rho_{2} \ll \rho_{3} \ll \cdots \ll
\rho_{\scriptscriptstyle{N} - 1},
\end{equation}

\noindent which represents a configuration where particle
3 is very far out off the area
spanned by particles 1 and 2, particle 4 is very far out
off the area spanned by
particles 1, 2 {\em and} 3, and so on.
In this limit, comparing eqns.(8) and (11) we have that

\begin{eqnarray}
\phi_{1j} & = & \phi_{2j} = \cdots = \phi_{j-1\,j} =
\phi_{j-1}, \nonumber \\
        j & = & 3, 4, \ldots, N.
\end{eqnarray}

\noindent The relation $\phi_{12}=\phi_{1}$ is always
trivially true,
independently of the cluster condition (13). As an example,
in the case $N=3$ the clustering condition implies

\begin{equation}
\phi_{13}=\phi_{23}=\phi_{2},
\end{equation}

\noindent while in the case of four anyons

\begin{eqnarray}
\phi_{13} & = & \phi_{23} = \phi_{2}, \nonumber \\
\phi_{14} & = & \phi_{24} = \phi_{34} = \phi_{3}.
\end{eqnarray}

The Lagrangian (12) in the limit (13) is a function only
of the polar Jacobi variables

\begin{equation}
L_{rel} = \frac{1}{2}\sum_{k=1}^{N-1}({\dot{\rho}}_{k}^{2} +
{\rho}_{k}^{2}{\dot{\phi}}_{k}^{2}) -
\frac{1}{2}\sum_{k=1}^{N-1}{\omega}^{2}{\rho}_{k}^{2} +
\alpha\hbar\sum_{k=1}^{N-1}k{\dot{\phi}}_{k},
\end{equation}

\noindent and the corresponding Hamiltonian reads

\begin{equation}
H_{rel} = \frac{1}{2}\sum_{k=1}^{N-1}(P_{\rho_{k}}^{2}
+ \frac{(P_{\phi_{k}} - k\alpha\hbar)^{2}}{\rho_{k}^{2}}
+ \omega^{2}\rho_{k}^{2}).
\end{equation}

\noindent This is the first main result of our analysis:
in the clustering limit, eq.(13), the $N$-anyon relative
Hamiltonian is separated in the sum of $N-1$
relative two-anyon oscillator Hamiltonians, with the
respective statistical parameters
$\alpha ,\, 2\alpha ,\, \ldots , \, (N-1)\alpha$,
and the associated Schroedinger problem can be solved
exactly.

The quantum
mechanics of the simplified hamiltonian system (18) and its
connections with
the complete solution of the many-anyon problem will not be
discussed here.
However, we want to remark that the situation described by
eq.(13) is only a subcase of a much more general condition
of separability of the Lagrangian (7). In fact, it is
always correct to
replace the azimuthal angles in (7) by the Jacobi polar
angles according to
the rule (14) as long as particle 3 winds up either
around both
particle 1 and particle 2 or around none of them,
but not around particle 1 or particle 2 alone; particle 4
winds up around particles 1, 2, and 3 together or around
none of them, but not
around one of them or two of them alone, and so on (see
Fig.2 for an illustration in the case $N=4$). This can
be understood reminding
that the topological term in the Lagrangian (7) can be
anything that describes
the correct windings of anyons, or, to put it
differently, anything
that belongs to the correct homotopy class of the problem: in
the above
described situation the winding orbits do not cross. On the
other hand, the
azimuthal and the Jacobi angles are related by the property
that the winding
number of a Jacobi trajectory encircling $m$ particles
is $m$ times the winding
number of an azimuthal trajectory for two particles
encircling each other; this
means that since in the two-anyon case each anyon moving
around the
other picks up a phase factor $2\pi\alpha\hbar$, in
the $N$-anyon case each anyon
going around the other $N-1$ anyons picks
up a phase factor $2(N-1)\pi\alpha\hbar$.
Therefore, the replacement of the Lagrangian (12)
by the Lagrangian (17) is
{\em exact} for non-crossing windings (see Fig.2).
The clustering condition (13) is then just
an example of the many possible configurations
leading to separability.

The energy spectrum for the quantum Hamiltonian (18) can
be derived exactly.
However, and this is our second main result,
it is also straightforward to derive it applying
the semi-classical
procedure described in the introduction. The result
is  the
generalization of eq.(6) to $N$ anyons:

\begin{equation}
E = \sum_{k=1}^{N-1}(2n_{k} + 1 +
|m_{k} - k\alpha|)\hbar\omega.
\end{equation}

\noindent The energy levels are linear in
the statistical parameter $\alpha$;
the radial quantum numbers $n_{k}$ should be
positive integers, while the
angular quantum numbers $m_{k}$ can be both
positive and negative. The correct
degeneracy can be obtained only from the full
quantum solution for the
Hamiltonian (18), and it will be
discussed elsewhere (see ref. [11]).
\vspace{0.8cm}

{\large \bf 4. Discussion and conclusions}

\vspace{0.5cm}

Formula (19) predicts that all energy levels
as function of $\alpha$ should
fall into different classes, each with
a positive and a negative slope.
For $N=4$ the levels fall into four different
classes with slopes $0,{\pm}2,{\pm}4,{\pm}6$. The
corresponding spectrum is drawn in Fig.3. Again, as
in the case of three anyons, it
reproduces the numerical results recently obtained
for the four-anyon system [10].
This is encouraging in view of a possible
close connection between
the solution of the simplified
model-Hamiltonian (18), and the solution
of the general anyon-gas model. Work is in
progress in that direction [11].

To conclude, we stress that the procedure
to derive the solvable Hamiltonian
(18) and the semi-classical spectrum (19)
does not depend on the details of
the external potential, and therefore it can
be easily applied to other related
models, e.g. to the motion of anyons
in a magnetic field, which has a great
physical relevance in connection with
the fractional quantum Hall effect.
The semi-classical picture for the anyon-gas
in external magnetic field
will be presented in a forthcoming paper [12].

The author is indebted to Jon Magne Leinaas, Pier
Alberto Marchetti and Mario Tonin for stimulating
discussions.
\newpage
{\begin{verse} {\Large {\bf References}} \\
\vspace{1.4cm}
[1] J. M. Leinaas and J. Myrheim, Nuovo Cimento
{\bf B 37}, 1 (1977). \\
\vspace{0.8cm}
[2] Y. Aharonov and D. Bohm, Phys. Rev. {\bf 115},
485 (1959). \\
\vspace{0.8cm}
[3] F. Wilczek, Phys. Rev. Lett. {\bf 48}, 1144 (1982). \\
\vspace{0.8cm}
[4] S.M. Girvin and R. Prange, The Quantum Hall Effect
(Springer, New York, 1990). \\
\vspace{0.8cm}
[5] F. Wilczek, Fractional Statistics and Anyon
Superconductivity (World Scientific, Singapore, 1991). \\
\vspace{0.8cm}
[6] Y.S. Wu, Phys. Rev. Lett. {\bf 53}, 111 (1984). \\
\vspace{0.8cm}
[7] G. Dunne, A. Lerda, S. Sciuto, and C.A. Trugenberger,
Nucl. Phys. {\bf B 370}, 601 (1992). \\
\vspace{0.8cm}
[8] F. Illuminati, F. Ravndal, and J.A. Ruud, Phys. Lett.
{\bf A 161}, 323 (1992). \\
\vspace{0.8cm}
[9] M. Sporre, J.J.M. Verbaarschot, and I. Zahed, Phys.
Rev. Lett. {\bf 67}, 1813 (1991). \\
\vspace{0.8cm}
[10] M.Sporre, J.J.M. Verbaarschot, and I. Zahed, SUNY
Preprint No. SUNY-NTG-91/40, 1991 (to be published). \\
\vspace{0.8cm}
[11] F. Illuminati, to be published. \\
\vspace{0.8cm}
[12] F. Illuminati, DFPD Preprint No. DFPD 92/TH/42, 1992
(to appear in Phys. Rev. Lett.).
\end{verse}}
\newpage
{\Large \bf Figure Captions}

\noindent {\bf Fig.1.} Semi-classical energy levels for the
three-anyon system in harmonic oscillator potential. Only
the levels which equal the numerical intercepts are reproduced.

\noindent {\bf Fig.2.} Different winding configurations for
the $4$-anyon problem.
In the examples A and B the relative dynamics is
exactly described either by the azimuthal or by the Jacobi
variables, and
the replacement of the Lagrangian (12) by the Lagrangian (17)
is correct.
This is not true in cases C and D.

\noindent {\bf Fig.3.} The semi-classical energy spectrum of
four anyons in harmonic oscillator potential.
\end{document}